\documentclass[pre,twocolumn,showpacs]{revtex4-1}%
\usepackage{amsfonts}
\usepackage{amsmath}
\usepackage{amssymb}
\usepackage{graphicx}%
\begin{document}

\title{Non-linear irreversible thermodynamics of single-molecule experiments}

\author{I. Santamar\'{\i}a-Holek$^a$\footnote{
E-mail: isholek.fc@gmail.com}, N. J. L\'opez-Alamilla$^a$, M. Hidalgo-Soria$^a$, A. P\'{e}rez-Madrid$^b$\footnote{E-mail: agustiperezmadrid@ub.edu}}

\affiliation{$^a$UMDI-Facultad de Ciencias,  Universidad Nacional Aut\'onoma de M\'exico Campus Juriquilla, Quer\'etaro 76230, M\'exico;
$^b$Departament de F\'isica Fonamental, Facultat de F\'isica, Universitat de Barcelona, Marti i Franques, 08028 Barcelona, Spain}



\begin{abstract}%
{Irreversible thermodynamics of single-molecule experiments subject to external constraining forces of a mechanical nature is presented. Extending Onsager's formalism to the non-linear case of systems under non-equilibrium external constraints, we are able to calculate the entropy production and the general non-linear kinetic equations for the variables involved. In particular, we analyze the case of RNA  stretching protocols obtaining critical oscillations 
between different configurational states when forced by external means to remain in the unstable 
region of its free-energy landscape, as observed in experiments. We also calculate 
the entropy produced during these hopping events, and show how resonant phenomena
in stretching experiments of single RNA macromolecules may arise. We also calculate the hopping rates using Kramer's approach obtaining a good comparison with experiments.
 }
\end{abstract}
\maketitle 

\section{Introduction}
Many mesoscopic systems of great interest ranging from biology, such as proteins, RNA or DNA molecules 
\cite{a,bustamante,thirumalai1,thirumalai1-b,thirumalai2,thirumalai3,thirumalai4,felix,la porta,FelixBJ2005}, 
to nano-engeneering, like rotaxanes and catenanes \cite{Balzani1997,Elbueno,Wurpel}, have similar dynamic behaviors due to their intrinsic properties and structure, as well as to their strong coupling with the surroundings. 
In particular, when they are subject to stressing conditions, like stretching forces or radiative interaction, the dynamics of these systems may present critical oscillations and stochastic resonance phenomena. 
Although well understood from the point of view of stochastic processes theory, the connection between  these dynamical behaviors and the thermodynamic properties of  these small systems and its coupling with 
changing surroundings is poorly understood \cite{landauer}. 

A better understanding of these peculiarities is of great importance in molecular biology, since it opens the 
possibility of correlating relevant physical-chemistry laws with simple models, thus allowing a transparent interpretation of experimental data and an agile prediction of different effects. Here, we establish this correlation through an irreversible thermodynamic description of biological small systems 
subject to external forces of a mechanical nature. This formalism is a powerful generalization of a previous analysis performed in Ref.  \cite{PRE2013} involving thermal constraints and permits one to calculate non-linear kinetic equations for the variables comprised and the entropy produced during the hopping events in both the critical oscillations and stochastic resonant phenomena. In this way, we can better distinguish  between reversible and 
irreversible behaviors of small systems and their coupling with the surroundings, through thermal and non-thermal fluctuations.

In order to highlight the ideas and simplify the presentation, we apply our formalism to analyze {\it single} RNA macromolecule stretching experiments which provide information on its energy landscape \cite{bustamante,thirumalai1,thirumalai1-b,thirumalai2,thirumalai3,thirumalai4,felix,la porta,cocco}. 
Specifically, the existence of a loop in the force-extension curve (FEC) implies that there is a range of tensions for which two stable configurations of the system are possible. We will show that  the transition between these two possible (folded/unfolded) configurations of RNA is concomitant with an irreversible generation of entropy by the molecule \cite{Rubi2006,Rubi2007}. 

Typically, the theoretical modeling of RNA macromolecule stretching experiments is performed by considering the composite system: RNA-macromolecule plus heat-bath, as an isolated system in thermal equilibrium. 
However, we propose here a novel approach to the problem of {\it single} macromolecule stretching experiments that differs from previous theoretical analysis because it is based on Onsager's non-equilibrium thermodynamics, 
specifically applied to the description of the time evolution of the state of the {\it single} macromolecule. 
This huge difference with respect to the mentioned previous descriptions, allows one to understand in a simpler and more consistent way the effect of fluctuations on both, the single macromolecule dynamics realizations as well as over the time averaged (or collective) properties that may be extracted from these experiments. Using the Onsager's theory to obtain the evolution equation for the extension of a {\it single} RNA molecule is a very general and simple form that does not needs to consider confusing or even inappropriate assumptions like ergodicity and equilibrium that are commonly used in the literature on this problem. In general terms, the idea is to calculate the entropy produced by the {\it single} RNA molecule during its transition between the folded and unfolded states, and to relate this with the corresponding free energy change between these configurations. This procedure is very rich since shows that the folding/unfolding process may not be viewed as an equilibrium situation and permits to deduce the stochastic non-linear kinetic equations that govern the hopping process for different conditions of the thermal bath and of the experimental devices.

It is convenient to mention here that some previous works have shown that entropy production may be associated to the process of transition between folded and unfolded states in {\it single} RNA macromolecule stretching experiments \cite{Rubi2006,Rubi2007}. However, they have been focused to show that non-equilibrium statistical arguments lead to the well established master equation formalism, which is the usual mathematical tool for the theoretical description of these systems \cite{cocco}. The important precedent here is that it was shown that the master equation formalism is compatible with the fact that entropy is produced during the hopping dynamics of the {\it single} RNA molecule stretching experiments.

The use of a master equation comes from the fact that it allows to determine the transition rates between the folded-unfolded states of the {\it single} RNA molecule, and suggests a chemical-like interpretation of the process. This description is sufficiently accurate from the quantitative point of view because it can be associated to average values in the thermal noise case. However, although non-Markovian master equations exist, using them to obtain practical results involving colored noises becomes a very difficult task that may substantially modify the obtained results and even their interpretation. Furthermore, it may result in confusion when used under the perspective of chemical kinetics, as we will discuss in detail. The evolution equation for the extension we derive using our generalization of Onsager's theory is of the Langevin type, and therefore can be easily generalized to more complex situations.

The article is organized as follows. First, we present our approach to calculate the entropy production 
of the system from which we obtain and analyze the non-linear stochastic kinetic equation accounting for the evolution of the variable determining the state of the system, the elongation of a RNA molecule. Afterwards, we solve numerically the equation showing four main results. We also calculate the entropy produced
during a single realization of hopping events. Then, we discuss the ensemble description that is widely used to reconstruct the free-energy landscapes of biological small-systems and discuss some important aspects in relation with the single molecule description. Finally,
we present our conclusions. 


\section{Dissipation in single molecule experiments, entropy production and kinetic equations}

We begin our approach to the problem by considering a {\it single} RNA molecule of length 
$ \langle x\rangle$ subject to a stretching force $\tau$ which can be controlled as an external parameter. 
According to thermodynamics \cite{Landau,Sokolov}, for the case of constant volume and temperature, 
the differential amount of work done by the surroundings on the molecule in order to change its elongation,
$dW \equiv \tau d \langle x\rangle$, satisfies the inequality
\begin{equation} \label{Landau1}
dW \geq dF,
\end{equation}%
where $dF$ is the differential of the Helmholtz free-energy. 
As it is well known, the minimum work  is performed in reversible processes for which the equality holds, 
implying that $dF=dW_{rev}=\tau d \langle x\rangle$. 

For unequilibrated irreversible processes, the same change of state comprises a larger work due to dissipation and the inequality
holds. Following  Ref. \cite{kondepudi}, it is easy to show that in the irreversible case the equality can be restored by adding
the entropy produced during the process, $Td_iS$. This leads to the relation
\begin{equation} \label{Landau2}
dW = dF + Td_{i}S.
\end{equation}%
This equation is very general since it is valid for variables determining the global state of the system 
(even if it is small) and may be used to determine the evolution equations of the system during the
unequilibrated process. From Eq. (\ref{Landau2}) it follows that 
the differential of the dissipated work is $dW_{dis}= Td_{i}S$.

In order to analyze the dissipative dynamics of the elongation of the RNA molecule, we have to notice 
that Eq. (\ref{Landau2}) gives the entropy produced by the system during 
the unequilibrated elongation process in terms of its free-energy change and the work done on the system
\begin{equation} \label{EntropProd1}
Td_{i}S= - dF  + \tau  d  \langle x\rangle .
\end{equation}%
From this equation the Onsager's theory may be generalized by assuming the existence of linear laws between fluxes and thermodynamic forces. As we will show, the flows are given by the time derivative of the independent variables, $d \langle x\rangle/dt$, whereas the forces are given as derivatives of the  free energy with respect to the independent variables: $dF/dx $, \cite{Groot-Mazur}. In this sense, the non-linear character of the resulting evolution equations comes from the fact that the free-energy is non-linear in the independent variables $x$, in contrast with original Onsager's theory, in which the forces are 
linear functions of the fluctuating variables.
Finally, it is important to mention  that in a previous work \cite{PRE2013} we discussed how equations of the form (\ref{EntropProd1}) can be derived based on  statistical mechanics and discuss the relation of these equations with the internal stability of small systems. As we have shown, a change  of convexity in the corresponding thermodynamic 
potential of a small system, irrespective of its nature, triggers irreversible processes that dissipate 
energy and, more importantly, establishes differences between the performed work and the 
corresponding free energy change as we discuss in the next subsections.

\subsection{Entropy production and stochastic kinetic equations}
In the general case when $dF \neq \tau d \langle x\rangle $,  
a dynamics is established in the system. This dynamics can be analyzed by computing the entropy production and
using the second law of thermodynamics within the framework of Onsager's irreversible thermodynamics \cite{kondepudi,Onsi-Machli}.
For this, we may divide Eq. (\ref{EntropProd1}) 
by $d t$ and take the limit $d t \rightarrow 0$ 
yielding the differential equation for the entropy production \cite{Groot-Mazur} 
\begin{equation}\label{EntropProd1-d}
T\frac{d_{i}S}{dt} = 
\left[-\frac{\partial F_{ \langle x\rangle}}{\partial  \langle x\rangle} +\tau \right]\frac{d  \langle x\rangle}{dt}.  
\end{equation}%
Since the second law imposes that the entropy production must be non-negative, $d_{i}S/dt>0$,
we may assume, without loss of generality, the relationship 
\begin{equation} \label{Dynamic-1} 
\frac{d \langle x\rangle}{dt}=  \gamma \left[-\frac{\partial F_{ \langle x\rangle}}{\partial  \langle x\rangle} +\tau \right], 
\end{equation}%
where $\gamma>0$ is an Onsager coefficient having dimensions of mobility. It is important to mention here that an equation similar to Eq. (\ref{Dynamic-1}) has been previously derived in Ref. \cite{Rubi2006} by using using mesoscopic non-equilibrium thermodynamics. Such a connection with the statistical Gibbs entropy is relevant for two reasons: {\it i}) It shows that during the transitions of the {\it single} molecule between configurations entropy is produced and, {\it ii}) at the same time, establishes a connection between this entropy production and the master equation formalism mentioned in the introduction, see Refs. \cite{cocco,Rubi2007}.
In addition, it must be stressed that the derivation given in Ref. \cite{Rubi2006} introduces an internal degree of freedom that allows one to describe in detail the distribution of tension and forces along the system length. When this internal variable is considered, a thermodynamic formulation of the problem needs the assumption of local equilibrium in the corresponding level of description. The equivalence of results shows  that the local equilibrium assumption remains valid for large gradients in small systems, like proteins \cite{BresmeT} and carbon nanotubes \cite{CNT} where the non-equilibrium thermal relaxation as been successfully studied. 

Equation (\ref{Dynamic-1}) is the overdamped dynamic equation for the average elongation of the
molecule and thus, meaning that at equilibrium (${d \langle x\rangle}/{dt}=0$) the external tension $\tau$ is compensated
by the internal tension: $\tau_{int} \equiv {\partial F_{ \langle x\rangle}}/{\partial  \langle x\rangle}$. 

In the case of small systems in contact with a heat bath, like in the {\it single} RNA-molecule experiments under consideration, thermal fluctuations should produce fluctuations of the instantaneous elongation of the molecule, $x(t)$, that in turn induce an unbalance of forces, $dF \neq \tau d  x $. This is the well known program of Onsager's irreversible thermodynamics \cite{Onsi-Machli,Sokolov}. Therefore, for small systems one has to formulate the stochastic version of Eq. (\ref{Dynamic-1}), that is given by the Langevin-like equation
\begin{equation} \label{Dynamic-stoch} 
\frac{d  x}{dt}= \gamma \left[\tau - \tau_{int}\right] + \xi (t),
\end{equation}%
where $\xi ({t})$ is an additive random Gaussian noise having zero mean and obeying the fluctuation-dissipation theorem: $\langle \xi ({t})\xi ({t}^{\prime })\rangle =2 D\delta ({t}-{t}^{\prime })$, with $D = k_BT \gamma$ the intensity of noise.  The derivation of Eq. (\ref{Dynamic-stoch}) can also be obtained by applying a fluctuating hydrodynamics scheme.

The key point here is that thermal fluctuations spontaneously modify the state of the system which develops a dynamics that depends on the stochastic nature of those fluctuations and on the internal and external constraints. In the case of stretching experiments with {\it single} RNA molecules that we are considering, two outputs are possible:
{\it i)} If the external tension is below or above of the bistable region of the force-extension curve, the elongation is single valued which means that these fluctuations should be Gaussian with respect to an average value of the elongation because the molecule is in equilibrium. {\it ii)} However, if the tension takes values in the portion of the force-extension curve with negative slope (corresponding to the convex part of the free energy) the system is intrinsically unstable and therefore thermal fluctuations should trigger the critical oscillations of the elongation of the molecule between the folded and unfolded states. The probability distribution associated to these fluctuations, although may take a stationary shape, is bimodal and it cannot be taken as an equilibrium distribution compatible with thermodynamic equilibrium. 
\begin{figure}[]
\begin{center}
\mbox{\resizebox*{7.5cm}{9.5cm}{\includegraphics{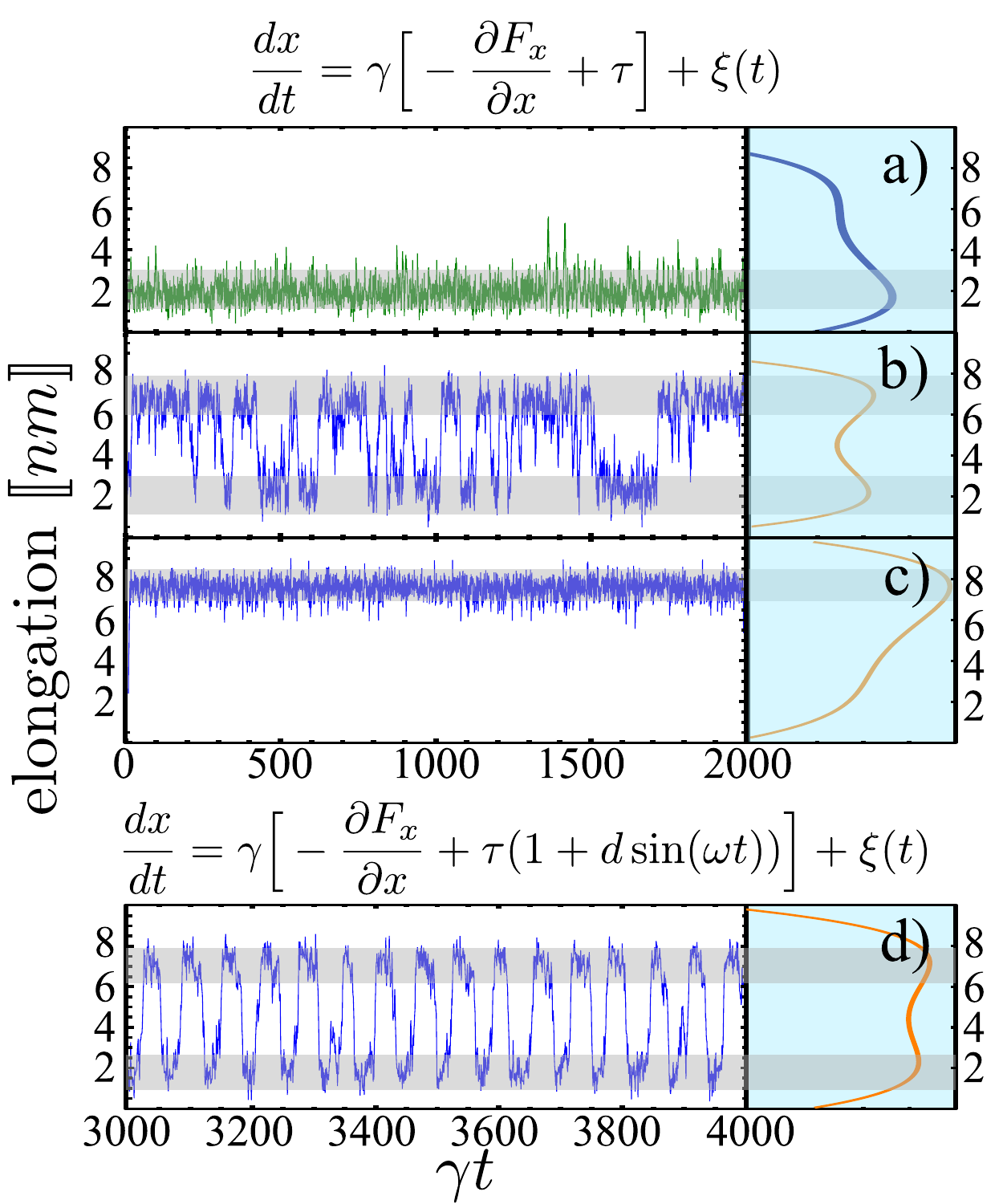}}} {} 
\end{center}
  \vspace{-6mm}
\caption{{\protect\footnotesize {(Color
  online) Elongation of RNA molecules subject to tension near the folding/unfolding states. It is also shown the corresponding shape of the free energy as obtained by fitting simulation data from \cite{thirumalai3}.
a) For low external tensions applied on the RNA molecule the fitting of the free energy implies that the secod term at the right hand side of Eq. (\ref{F}) is: $\tau_{int}^{(c)}-\tau=-4.72$pN, where the absolute minimum of the free energy corresponds to the folded configuration. b) For external tensions in the range of the unstable region of the force-extension curve, the free energy is bistable and the effective tension takes the value  $\tau_{int}^{(c)}-\tau=-5.24$pN.
c) For large enough tensions, $\tau_{int}^{(c)}-\tau=-6.42$pN, the absolute minimum of the free energy 
corresponds to the unfolded configuration. d) Effect of a periodically driving force $\tau(t)$ on the hopping dynamics of the molecule. The parameters of the oscillation are $\omega=\gamma/10$ and} $\tau_{int}^{(c)}-\tau \simeq -5{.}32$pN.}}
\vspace{-6mm}
  \label{h}
\end{figure}

According to the experimental and simulation results reported in \cite{thirumalai1}, for tensions around
15pN, the intrinsic free-energy of 
the RNA molecule is bistable as a function of the elongation $x$. 
A series expansion of $F_{\langle x\rangle}$ around the critical value of the elongation 
$\langle x\rangle = x_c +x $, yields the Landau-De Gennes-like model
\begin{equation} \label{F} 
F_{x} \simeq F_c + \tau_{int}^{(c)} x +\kappa_{0} x^2 - \kappa_{1} x^3 + \kappa_{2} x^4,
\end{equation}
where the coefficients $\kappa_{0}$, $\kappa_{1}$ and $\kappa_{2}$ 
are considered positive and $\tau_{int}^{(c)}$ is the tension of the RNA molecule at the critical value $x_c$
of the elongation. The internal tension is therefore given by 
$
\tau_{int}  \simeq \tau_{int}^{(c)} + 2 \kappa_{0}  x - 3 \kappa_{1} x^2 + 4 \kappa_2 x^3.
$
Upon substituting the last expression into Eq. (\ref{Dynamic-stoch}), an effective free-energy 
can in turn be defined that incorporates the effect of the externally imposed tension $\tau$, that is
\begin{equation} \label{F} 
F_{e} \simeq F_c + \left(\tau_{int}^{(c)}-\tau\right) x +\kappa_{0}  x^2 - \kappa_{1} x^3 + \kappa_{2} x^4,
\end{equation}
The external tension tilts the internal free-energy landscape $F_{x}$ allowing transitions between 
the two possible configurations, folded and unfolded. 
The equation for  the average elongation is now
\begin{equation} \label{x con Fef} 
\frac{d\langle x\rangle}{dt}=  - \gamma\frac{\partial F_{e}}{\partial \langle x\rangle}.
\end{equation} 
The effective free-energy  is reported in experiments and simulations from which the values of
the parameters and important biological information 
can be obtained \cite{bustamante,thirumalai1,thirumalai2,thirumalai3,thirumalai4,Keller}. 

The numerical solutions of Eq. (\ref{Dynamic-stoch}) for the free-energies considered were obtained 
using the Euler-Maruyama scheme and  presented in Figures 1 and 2. Critical oscillations appear when the external and  internal tensions balance each other out in such a way that the effective free energy 
$ F_{e}$ has two minima with similar energies. For sufficiently low or large external tensions, the free energy has a dominant minimum corresponding to the folded and unfolded states, respectively [see Figures 1a)-c)].
The noise induces critical oscillations between the two stable states separated by the free-energy barrier.
The stochastic process associated with this oscillation of the molecule's elongation is not 
a Gaussian white noise, as should occur for equilibrium fluctuations. On the contrary, it resembles a dichotomous stochastic process having a finite correlation time due to the intermittency which causes hopping events between
both metastable states of the system. As it is well known, this  intermittency, characterized by non-Gaussian distributions, is one of the characteristics of non-equilibrated processes \cite{felix2,Agustin}. 

It is also important to emphasize here that the critical oscillations associated to the hopping events cannot be
understood as an example of an equilibrated chemical reaction 
\cite{bustamante,thirumalai4}. The first aspect of the problem is to consider that unlike the concentrations of chemical species in equilibrium chemical reactions, a {\it single} molecule can never be in both states at the same time (coexistence). The second aspect is that the thermodynamics of chemical reactions involves the concept of concentration, which in statistical language entails the assumption of an ensemble of systems. For dilute compounds, concentration and probability are related concepts. Thus, any chemical information and chemical kinetics description implicitly corresponds to an ensemble of systems, and not to individual ones. Our contention here, is that during a single experiment, each molecule cannot be in equilibrium because it jumps alternatively between the folded and unfolded states; its elongation is an oscillatory function of time. These hopping events imply a finite change of length of the molecule at a finite rate and thus cannot be seen as an adiabatic hopping process. It may happen, however, that in the case of a collectivity of RNA molecules the number of molecules passing from the folded to the unfolded states and viceversa, compensate each other, thus leading to the well known definition of the equilibrium constant in terms of the average reaction rates of a chemical reaction.  The crucial point here is very well understood in statistical physics: Although the ensemble may be in a stationary state (there is not clear evidence of a thermodynamic equilibrium here), each system (molecule) develops a dynamics, that is, a non-equilibrium process that should satisfy the well known fluctuation-dissipation theorem.
\begin{figure}[]
\begin{center}
\mbox{\resizebox*{7.5cm}{4.5cm}{\includegraphics{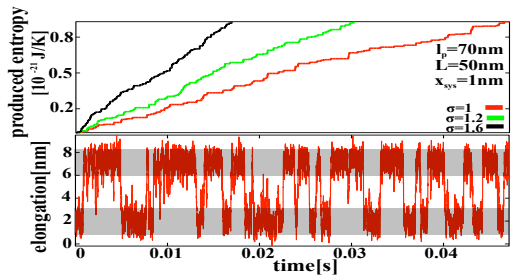}}} {} 
\end{center}
  \vspace{-0mm}
\caption{{\protect\footnotesize {(Color online) The entropy produced as a function of time (lines) 
by integrating Eq. (\ref{EntropProd1-d}) for the single noise realization shown. Different
intensities of the noise term $\sigma = k_BT \gamma$ where considered. Increasing noise
intensity (larger $\sigma$) implies larger entropy produced because hopping events are
promoted. $l_p$ is the persistence length of the handles, $L$ the length of the handles and $x_{sys}$ the dimensions of the hairpin according to Ref. \cite{thirumalai2}.}}}
\vspace{-6mm}
\label{g}
\end{figure}

The last result indicated in Figure 1d) shows an example of the effect of a periodically time dependent force on the critical oscillations. Depending on the intensity of this oscillation, stochastic resonance and a more controlled jumps between folded and unfolded states may be obtained, thus allowing to infer more regular shapes for the free energy. 
It should be stressed that this oscillating force is a toy model of existent feedback mechanisms used in the experimental devices  to maintain a constant tension when spontaneous oscillations (folded and unfolded configurations) take place. Many details of this interesting aspect of experimental setups and their theoretical modeling are well discussed in Refs. \cite{Felix-Fedback1,Felix-Fedback2} for the so-called passive or constant force modes. Here, we assumed a time dependent tension imposed on the system: $\tau(t) = \tau [1+d \sin(\omega t)]$, with $d$ an amplitude factor and  $\omega$ the frequency, with the aim to emphasize that this kind of mechanisms may modify the response of the system as shown in Figure \ref{h}d),
leading to stochastic resonance events \cite{felix}. The results that can be obtained in this case have a direct influence on the probability distributions (or histograms) obtained from simulations and/or experiments, and therefore may influence some aspects of the determination of the free energy profiles.

\subsection{Entropy produced by a single molecule in a single realization}
For a single realization of the noise and, therefore, for a single sequence of hopping events of one RNA molecule, it is possible 
to calculate the entropy produced by numerically integrating Eq. (\ref{EntropProd1-d}). 
Furthermore, a general
analytic expression for the entropy produced can be obtained by realizing that the hopping
events can be modeled with sum of Heaviside functions having an stochastic distribution
of transition times $t_i$: 
\begin{equation}\label{x-comb}
x(t) \simeq \sum_i \Delta x \Theta[t-t_i], 
\end{equation}
with $\Delta x$ the elongation
change between folded and unfolded states. The time derivative of this function is known as
the comb function $\text{III}(t-t_i) \equiv  \sum_i \delta(t-t_i)$ 
\begin{equation}\label{x-comb}
\frac{d x}{dt} = \Delta x\,\text{III}(t-t_i).
\end{equation}
Using this result into \ref{EntropProd1-d} and integrating over time one obtains that
the entropy produced during a given elapsed time is
\begin{equation}\label{EntropProd1-f}
{\Delta_{i}S} = \frac{\Delta x}{T}\sum_i \left[\tau-\tau_{int}(t_i)\right].
\end{equation}%
For the realization of the noise shown in Figure 2, it is also shown the entropy
produced in terms of time. The average behavior resembles a stair-like increase due to
the hopping events between the two possible states. Thermal fluctuations in the
unfolded and folded states of the dynamics where subtracted in order to keep
only the hopping events.

\section{Ensemble description}

The results obtained in the previous subsection refer to a {\it single} RNA molecule. However, the stochastic nature of the dynamics of the system allows one to introduce an interesting aspect of the problem that is associated with the {\it collective} behavior of an ensemble of equivalent systems. This collective behavior can be analyzed in terms of a  probability density whose time evolution is concomitant with the Langevin-like equation Eq. (\ref{Dynamic-stoch}).

For the present case with additive white noise, the Fokker-Planck equation for the probability 
distribution of $x$, $P(x,t)$, associated to Eq. (\ref{Dynamic-stoch}) is
\begin{equation}
\frac{\partial P}{\partial t} =\gamma \frac{\partial }{\partial x}\left[ \left(\tau_{int} - \tau\right)  P+ k_BT \frac{\partial P}{\partial x }\right].  \label{Dynamic-FP}
\end{equation}%
For constant external tensions, it is well known the stationary solution of this equation
\begin{equation}\label{Eq-Sol-FP}
P_{eq}(x) =  Q^{-1} {e^{-\left[F(x)-\tau x\right]/k_BT}}=  Q^{-1} {e^{-F_{e}(x)/k_BT}},  
\end{equation}
with $Q =\int^\infty_\infty e^{-\left[F(x^\prime)-\tau x^\prime\right]/k_BT }dx^\prime$.
This expression predicts a distribution function that may be  bimodal  depending on the balance 
between the external and internal tensions, \cite{Keller}. For example, for an externally applied tension of 15pN [see Ref. \cite{thirumalai3}], the value of the fitting parameter of the linear term of the free energy is: $\tau_{in}^{(c)}-\tau = - 5.22$pN. In this case, the effective free energy $\Delta F_{e}$ has two well defined minima and it follows that the internal tension of the RNA molecule during the experiments was $\tau_{in}^{(c)} \simeq 9.78$pN. This is the tension associated to the handles at the critical elongation value, whereas the remaining $5.22$pN are the apparent internal tension of the hairpin during the unfolding/folding process.

In Figure 3 we show the spatio-temporal numerical solution of Eq. (\ref{Dynamic-FP}) for different shapes of the potential (also shown). For constant external tensions, stationary distributions are obtained that perfectly agree
with Eq. (\ref{Eq-Sol-FP}). Their shape depends on the shape of the free energy and can be
bimodal or single-modal depending on the value of the external tension.
\begin{figure}[]
\begin{center}
\mbox{\resizebox*{7.5cm}{4.5cm}{\includegraphics{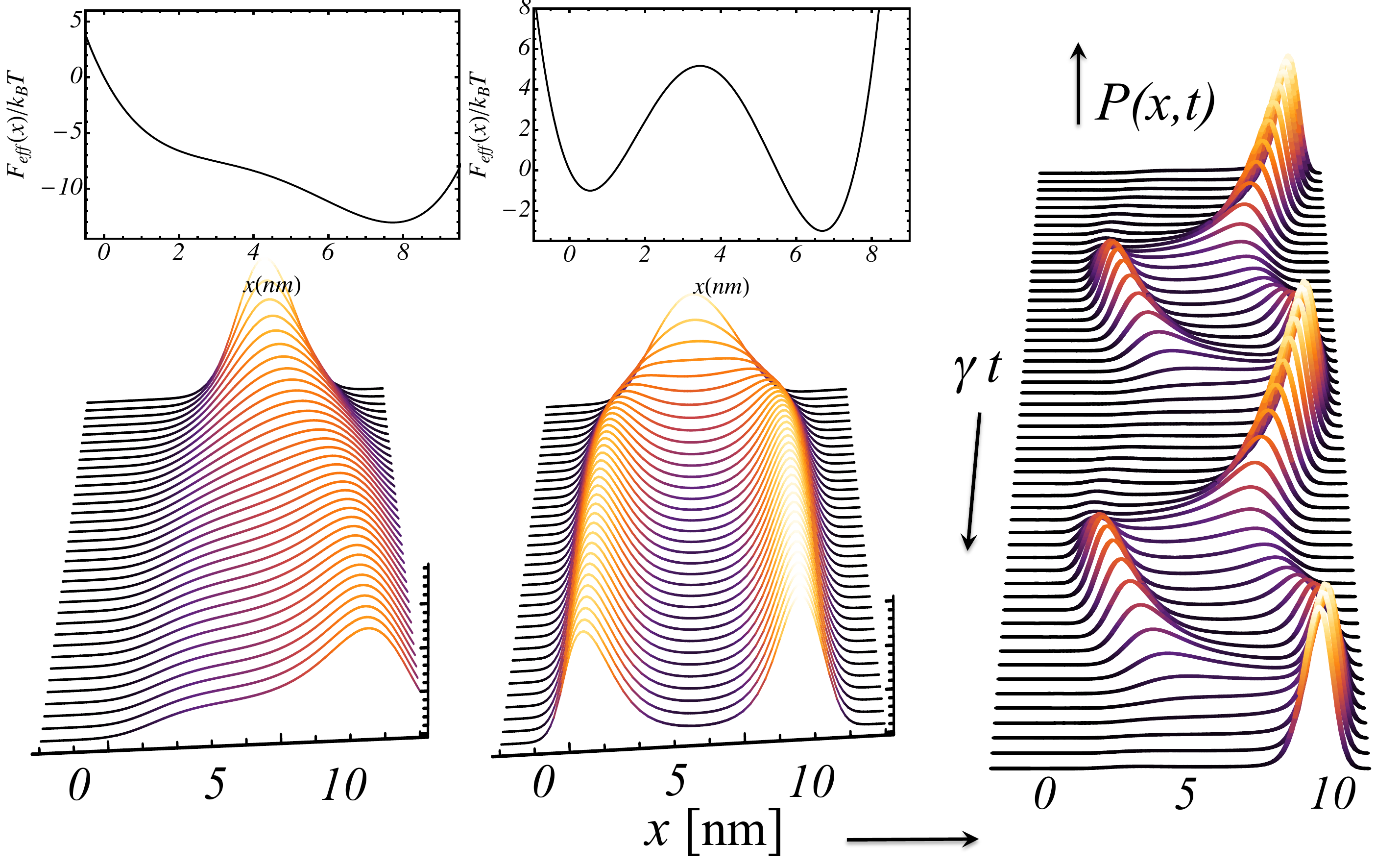}}} {} 
\end{center}
  \vspace{-0mm}
\caption{{\protect\footnotesize {(Color online) Three probability distributions as a function of elongation for different times obtained numerically from Eq. (\ref{Dynamic-FP}). a) Two probability distributions (modal non-Gaussian and bimodal) from the initial stage to the stationary regime in the case of a constant external tension.  b) Two oscillatory behaviors of the bimodal distribution in the presence of a periodic external force, the alternative increase of probability in folded and unfolded states and the oscillation of the maximum corresponding to the unfolded state.}}}
\vspace{-6mm}
\label{g}
\end{figure}

However, in the case of stochastic resonance that involves the external time-dependent tension, 
$\tau(t) = \tau [1+d \sin(\omega t)]$, the resulting probability densities are time
dependent with no stationary limiting behavior. Oscillations of each maxima of the distributions are
clear and its amplitude depend on the amplitude of the externally applied force. For bimodal distributions
there are two oscillations coupled, that of the maxima with respect to an average value and
that of the alternating change of the populations between the two possible states, folded and unfolded,
see Figure 3.
For single-modal distributions, only the first oscillation takes place. This is an important feature since if time averages are not taken over sufficiently large time windows, they can lead to a 
incorrect determination of a bimodal distribution and therefore to an incorrect determination 
of the free energy landscape.

The probability distributions obtained in this subsection show more readily the main features of the nature of the elongation's fluctuations we already discussed in the previous subsection. For instance, if the external tension is above but near the bistable region of the force-extension curve, then the elongation of the RNA molecule is single valued (the free energy has only one minimum) but its distribution shows a large asymmetry that indicates a non-Gaussian distribution for the fluctuations of the molecule's elongation. Such a situation is shown in the left panel of Figure 3 and is in good agreement with the histogram of Figure 1c) (not shown).  Hence, at the edge of the bistable zone, these results clearly show a break of the classical condition for equilibrium fluctuations. 

This effect is more dramatic when considering the case when the tension takes values in the portion of the force-extension curve with negative slope. The distribution of these fluctuations is stationary and bimodal. The main implication of this result is that the collective behavior reaches a stationary state characterized by a non-Gaussian distribution showing two peaks. These two peaks imply two most probable elongations of the molecule that do not correspond with the average elongation of the system, just at the center of the distribution. Hence, it is much less probable to find the system with the average elongation than in an extreme configuration. 
This fact contradicts the statistical condition of equilibrium which is consistent with the condition of thermodynamic equilibrium: the average value of the elongation in the single peaked probability distribution is equal to the most probable value of the same variable \cite{Landau}. 

\subsection{Transition rates and mean first passage times}

Two important quantities characterizing the hopping dynamics of the RNA molecule between the folded an unfolded states 
are the transition rates and their inverse, the mean first passage times. In the present case, the 
transition rates can be calculated using the Fokker-Planck equation (\ref{Dynamic-FP}) in the 
non-equilibrium stationary case \cite{Risken}. As usual, the Fokker-Planck equation 
(\ref{Dynamic-FP}) is written in the form 
$\frac{\partial P}{\partial t} = - \frac{\partial }{\partial x} j $,
in which the current $j$ is defined as
\begin{equation}
j = - \left[\frac{D}{k_BT}\frac{dF_{e}}{dx}  P+ D \frac{\partial P}{\partial x }\right],
 \label{stationary-j}
\end{equation}%
where $j$ is assumed to be a constant different from zero and $D=k_BT\gamma$. If we define $x_{f}$ as the elongation of the folded state, then the complete solution of this equation in the interval $[x_{f},x]$ is given by
\begin{eqnarray}
P(x)= P(x_{f}) e^{-\left[F_{e}(x)-F_{e}(x_{f})\right]/k_BT} \,\,\,\,\,\,\,\,\,\,\,\,\,\,\,\,\,\,\,\, \\
\,\,\,\,\,\,\,\,\,\,\,\,\,\,\,\,\,\,\,\, - \frac{j}{D}e^{-F_{e}(x)/k_BT} \int_{x_{f}}^x {e^{F_{e}(z)/k_BT}dz}. \nonumber
 \label{solucion-stationary-j1}
\end{eqnarray}%
Solving Eq. (16) for $j$ and evaluating the result in $x_{u}$, the elongation of the unfolded state, we obtain
\begin{equation}
j = D \frac{P(x_{f}) e^{F_{e}(x_{f})/k_BT}-P(x_{u}) e^{F_{e}(x_{u})/k_BT}}{\int_{x_{f}}^{x_{u}} {e^{F_{e}(z)/k_BT}dz}}= j_f - j_u,
 \label{solucion-stationary-j2}
\end{equation}%
where $j_f$ and $j_u$ are the forward and backward currents, respectively. 
For sufficiently large energy barriers, $F_{e,max}>>k_BT$, one may assume fast local equilibration of the system in each well and write $P(x_{f})$ and $P(x_{u})$ in terms of Eq. (\ref{Eq-Sol-FP}). In addition, the integral in the denominator of Eq. (\ref{solucion-stationary-j2}) can be calculated by extending the limits of the integration from $[-\infty, \infty]$ and assuming that $F_e(x) \simeq F_{e}(x_{c}) - \frac{1}{2} F_{e,c}^{''}(x-x_{c})^2 $, where $F_{e,c}^{''}$ is the second derivative of $F_e(x)$ evaluated at $x_c$, the local maximum of the free energy.  These operations give the following expressions for the forward and backward currents
\begin{eqnarray}
j_f = D Q_f^{-1} \left[\frac{F_{e,c}^{''}}{2\pi k_BT}\right]^{1/2} e^{-F_{e}(x_{c})/k_BT}, \nonumber \\ 
j_u = D Q_u^{-1} \left[\frac{F_{e,c}^{''}}{2\pi k_BT}\right]^{1/2} e^{-F_{e}(x_{c})/k_BT},
 \label{ja y jb}
\end{eqnarray}%
where $Q_f$ and $Q_u$ are the normalization factors of the folded and unfolded states, respectively.
Finally, the transition rates are defined as $k_f \equiv j_f/n_f$ and  $k_u \equiv j_u/n_u$ where
 \begin{eqnarray}
n_f = Q_f^{-1}  \left[\frac{F_{e,f}^{''}}{2\pi k_BT}\right]^{-1/2} e^{-F_{e}(x_{f})/k_BT}, \nonumber \\
n_u =  Q_u^{-1} \left[\frac{F_{e,u}^{''}}{2\pi k_BT}\right]^{-1/2} e^{-F_{e}(x_{u})/k_BT}.
 \label{na y nb}
\end{eqnarray}%
Here, we evaluated the normalization factors by assuming the usual quadratic approximation of the free energy around each well, for instance: $F_e(x) \simeq F_{e}(x_{f}) + \frac{1}{2} F_{e,f}^{''}(x-x_{f})^2 $. Then, after substituting into the definition of the rates we obtain the final expressions \cite{Risken,ISH}
\begin{eqnarray}
k_f = \gamma \frac{\sqrt{F_{e,f}^{''}F_{e,c}^{''}}}{2\pi} e^{-\left[\frac{F_{e}(x_{c})-F_{e}(x_{f})}{k_BT}\right]}, \nonumber\\
k_u =\gamma \frac{\sqrt{F_{e,u}^{''}F_{e,c}^{''}}}{2\pi } e^{-\left[\frac{F_{e}(x_{c})-F_{e}(x_{u})}{k_BT}\right]}.
 \label{ka y kb}
\end{eqnarray}%
These quantities may be evaluated explicitly after fitting the free energy profiles reported in the literature \cite{thirumalai3}. Table 1 shows the values of the transition rates and the corresponding transition times evaluated for some cases. All these results fall within the same order of magnitude of the transition times reported in \cite{thirumalai3}.

\begin{widetext}
\[
\begin{tabular}{|c|c|c|c|c|c|c|c|}
\hline 
Case & $x_{F}[nm]$ & $x_{U}[nm]$ & $x_{T}[nm]$ & $\tau_{f}[s]$ & $k_{f}[{1}/{s}]$ & $\tau_{u}[s]$ & 
$k_{u}[{1}/{s}]$ \\ 
\hline 
1 & $1.93$ & $7.28$ & $3.71$ & $3.9 \times 10^{-4}$ &$2564.10$   & $5.9 \times 10^{-4}$ & $1694.91$ \\ 
\hline 
2 & $1.84$ & $7.20$ & $3.85$ & $3.5 \times 10^{-4}$ & $2857.14$ & $4.4 \times 10^{-4}$ & $2272.72$ \\ 
\hline 
3 & $1.96$ & $7.11$ & $3.92$ & $4.6 \times 10^{-4}$ & $2173.91$ & $4.5 \times 10^{-4}$ & $2222.2$ \\ 
\hline 
4 & $2.13$ & $6.29$ & $3.6$ & $1.7 \times 10^{-4}$ & $5882.35$ & $11 \times 10^{-4}$ & $909.09$ \\ 
\hline 
5 & $2.00$ & $7.41$ & $3.91$ & $3.9 \times 10^{-4}$ & $2564.10$ & $5.8 \times 10^{-4}$ & $1724.13$ \\ 
\hline 
6 & $2.04$ & $6.90$ & $3.89$ & $4.9 \times 10^{-4}$ & $2040.81$ & $6.1 \times 10^{-4}$ & $1639.34$ \\ 
\hline 
\end{tabular}
\]
Table I. Mean transition times and rates from the folded  to unfolded states obtained from Eqs. (\ref{ka y kb}) for the fitted energy profiles reported in \cite{thirumalai3}.
\end{widetext}

\section*{Conclusions}
As a summary, in this work we have proposed an irreversible thermodynamics of single-molecule experiments
subject to external constraining forces of a mechanical nature. Our work is a powerful extension of Onsager's formalism to the case of systems under non-equilibrium external constraints. Using this formalism we were able to
calculate the entropy production of a small system in contact with a bath and to derive the general non-linear kinetic equations for the variables involved. 
A more fundamental justification of our formalism can be given 
by postulating the existence of a partition function of the biological small-system in the presence of external constraints 
and the validity of the  thermodynamic stability criterion. Using these postulates,
an analytic continuation of the free-energy into the unstable region allows one
to relate the free-energy barrier with the entropy produced by the single system
during its time evolution. 

As an illustration, we have analyzed the folding-unfolding dynamics of RNA molecules
subject to tension. We  showed that a single macromolecule evolves far from equilibrium executing 
critical oscillations when it is forced by the external constraints to remain in the unstable region 
of its free energy. Since this free energy contains anharmonic contributions, these
critical oscillations are not a simple Gaussian process. This is a manifestation of the
unequilibrated nature of this process for which the entropy produced during hopping events was calculated.

The approach we propose uses global thermodynamic variables for the single small-system
in similar way as Onsager's non-equilibrium thermodynamic theory. This fact
enables us to derive non-linear  differential kinetic equations governing the dynamics of the
system in both, the deterministic and stochastic cases \cite{kondepudi,Onsi-Machli}. 
Unlike Onsager's original theory, we have shown the validity of this description for
systems far from equilibrium (non-linear regression in the presence of external forces) 
and not only near equilibrium  without external forces (linear regression of spontaneous fluctuations). 
This opens many possibilities for the theoretical analysis and prediction
of the behavior of single molecule experiments.

\vspace{3mm}  



\section*{Acknowledgements}
This work was supported by UNAM DGAPA Grant No. IN113415 and by the MICINN
of Spain under grant No. FIS2011-22603. APM  acknowledges financial support by the 
Academic mobility program of the National Autonomous University of Mexico and of 
UMDI-Faculty of Sciences UNAM Campus Juriquilla where this work was done during 
a sabbatical leave.

\end{document}